\documentclass[aps,prl,twocolumn,epsfig,amsfonts,amsmath,amstex,floatfix]{revtex4}

\usepackage{bbold}
\usepackage{color}

\usepackage{amsfonts,amsmath}
\usepackage{epsfig,amsmath}
\usepackage{graphicx}
\usepackage{dcolumn}
\usepackage{bm}
\usepackage{bbold}

\newcommand{\beq}{\begin{equation}}
\newcommand{\eeq}{\end{equation}}
\newcommand{\beqa}{\begin{eqnarray}}
\newcommand{\eeqa}{\end{eqnarray}}

\newcommand{\la}{\langle} 
\newcommand{\ra}{\rangle}

\def\jpb#1{{ J.\ Phys.\ B} {\bf#1}}

\def\lphys#1{{ Laser\ Phys.} {\bf#1}}

\def\nphy#1{{Nature\ Phys.} {\bf#1}}
\def\natphot#1{{ Nat.\ Phot.} {\bf#1}}
\def\njp#1{{ New\ J.\ Phys.} {\bf#1}}

\def\oc#1{{ Opt.\ Commun.} {\bf#1}}
\def\ol#1{{ Opt.\ Lett.} {\bf#1}}

\def\pla#1{{ Phys.\ Lett. A\/} {\bf#1}}

\def\pra#1{{ Phys.\ Rev. A\/} {\bf#1}}

\def\pre#1{{ Phys.\ Rev. E\/} {\bf#1}}
\def\prl#1{{ Phys.\ Rev.\ Lett.} {\bf#1}}

\def\rmp#1{{ Rev. \ Mod. \ Phys.} {\bf#1}}

\begin{document}

\title{Bridging Coherence Optics and Classical Mechanics -- \\ A Universal Light Polarization-Entanglement Complementary Relation}
\author{Xiao-Feng Qian}
\email{xqian6@stevens.edu}
\author{Misagh Izadi} 
\affiliation{Center for Quantum Science and Engineering, and Department of Physics, Stevens Institute of Technology, Hoboken, New Jersey 07030, USA}

\date{\today }

\begin{abstract}
While optics and mechanics are two distinct branches of physics, they are connected. It is well known that geometrical/ray treatment of light has direct analogies to mechanical descriptions of particle motion. However, connections between coherence wave optics and classical mechanics are rarely reported. Here we explore links of the two for an arbitrary light field by performing a quantitative analysis of two optical coherence properties: polarization and entanglement (implied by a wave picture of light due to Huygens and Fresnel). A universal complementary identity relation is obtained. More surprisingly, optical polarization, entanglement, and their identity relation are shown to be quantitatively associated with mechanical concepts of center of mass and moment of inertia through the Huygens-Steiner theorem for rigid body rotation. The obtained result bridges coherence wave optics and classical mechanics through the two theories of Huygens. 
\end{abstract}

\maketitle
{\noindent \bf Introduction}: As one of the greatest scientists of all time, Christiaan Huygens had made groundbreaking contributions to many branches of natural science with two best known fields in physics: optics and mechanics \cite{Aldersey-Williams}. Huygens is considered as the starting point of systematic wave explanation of light in the 1670s \cite{Ziggelaar80} and the Huygens-Fresnel principle \cite{Fresnel1818} was the basis for the advancement of physical optics, describing coherence phenomena of light including interference, diffraction, polarization, etc.~\cite{BornWolf}, as well as the recently recognized property of vector-space entanglement \cite{Spreeuw1998FP, Luis2009OC, Simon2010PRL, Borges2010PRA, Qian2011OL, Kagalwala2013NP, Toppel2014NJP, Zela2014PRA, Forbes2017Nat, Qian2017opn, Forbes2019PIO, Goldberg2021AOP} (a direct consequence of a multi-degree-of-freedom amplitude wave theory \cite{Qian2017opn}). On a completely different subject, through the study of pendulum oscillation that led to his invention of the first pendulum clock, Huygens also made pivotal contributions to the development of fundamental mechanical concepts of center of mass (COM) and moment of inertia (MOI) describing rigid body motions, leading to the well known Huygens-Steiner theorem (also called the parallel-axis theorem) \cite{Mach1919}. Except that both owe to the contributions of Huygens, almost no links of the two theories is ever explored or even anticipated due to their apparent distinctions. To fill this gap, here we provide an approach that connects the two subjects through the analysis of two optical coherence properties: {\em polarization} and {\em entanglement.}

As one of the earliest discovered fundamental features of light, polarization was only gradually better understood along with the slow recognition of the light's wave nature \cite{BornWolf,Stokes1852}, and it is conventionally understood as the directional property, or degree of freedom (DOF), of light (electromagnetic) wave oscillation. Recently, it has been further shown that polarization coherence needs at least one additional DOF to be fully characterized  \cite{Qian2011OL, Eberly2016LP}. This allows the discussion of its connection to another two-DOF property, i.e., entanglement \cite{Qian2016PRL}. Here we carry out a systematic analysis of both polarization (${\cal P}$) and entanglement (${\cal K}$) for a generic light field and obtain a universal complementary relation ${\cal P}^2+{\cal K}^2=1$ regardless of the dimensionality.  


On the other hand, attempts of geometric understanding of entanglement have been made in various contexts \cite{Bengtsson-Zyczkowski, Walter2013Sci, Alonso2016PRA, Qian2018NJP, Yang2022PRA}. With a geometric mapping of optical coherence parameters to a point-mass system, we further establish a surprising quantitative relation between the obtained universal optical polarization-entanglement complementary identity and the rigid body Huygens-Steiner theorem through specific mechanical concepts of COM and MOI. Our method and results suggest a new way of investigating quantitative and conceptual connections between coherence optics and mechanics.\\


{\noindent \bf Polarization-Entanglement Complementary Relation}: We start with the most general arbitrary light field that can be written in the Dirac notation \cite{Qian2011OL} as
\beq \label{Main3D-field}
|E\rangle=|x\rangle |E_x\rangle+|y\rangle |E_y\rangle+|z\rangle |E_z\rangle,
\eeq
where $|x\ra$, $|y\ra$, $|y\ra$ represent the three polarization (field oscillation) vector directions, and $|E_x\ra$, $|E_y\ra$, $|E_z\ra$ stand for corresponding field amplitudes describing the remaining degrees of freedom that live in respective vector spaces (temporal mode, spatial mode, etc.). When normalized by the total light intensity $I=\la E_x|E_x\ra+\la E_y|E_y\ra+\la E_z|E_z\ra$, it becomes
\beq \label{3Dnorm-field}
|e \rangle =\alpha |x \rangle |e_x \rangle + \beta|y \rangle |e_y \rangle + \gamma |z \rangle |e_z \rangle ,
\eeq
where $\alpha$, $\beta$ and $\gamma$ are real normalized coefficients defined as $\alpha = \sqrt{\langle E_x |E_x \rangle/I}$, $\beta =\sqrt{ \langle E_y |E_y \rangle/I}$ and $\gamma =\sqrt{ \langle E_z |E_z \rangle/I} $ with $\alpha ^2 + \beta^2 + \gamma^2=1$. The normalized amplitude vectors are defined as $|e_i\ra =E_i/ \sqrt{\la E_i |E_i\ra }$ with $i = x , y, z$. For the generic three-dimensional (3D) field, the cross correlation among amplitude components $|e_{x,y,z}\ra$ can be arbitrary and most generally described by complex values $\delta_1=\langle e_x |e_y \rangle $, $\delta_2=\langle e_x |e_z \rangle$, and $\delta_3=\langle e_y |e_z \rangle$ respectively.

The 3D polarization coherence of the light field can be characterized by the $3\times 3$ coherence matrix \cite{Setala2002PRE, Sheppard2012OL, Gil2004, Korotkova2017OL, Alonso2020arXiv},
which can be decomposed into nine Gell-Mann matrices \cite{Fano1957RMP}, and is obtained as
\beq \label{3Dnorm-matrix}
{\cal W}_{\rm 3D}=
\begin{bmatrix} 
\alpha^2 & \alpha\beta\delta_1 & \alpha\gamma\delta_2 \\
\alpha\beta\delta^*_1 &
\beta^2 & \beta\gamma\delta_3 \\
\alpha\gamma\delta^*_2 & \beta\gamma\delta_3^* & \gamma^2\\ 
\end{bmatrix}.
\eeq

Due to the fact that this $3\times 3$ matrix cannot be uniquely decomposed into the summation of a completely polarized matrix and a completely unpolarized partner \cite{BornWolf}, different measures of degrees of polarization have been proposed based on different interpretations of complete unpolarization \cite{Alonso2020arXiv}. To have a systematic analysis for arbitrary dimensions (e.g., the 2D arbitrary beam, the 3D arbitrary field, etc.), here we adopt the point of view that complete 3D unpolarization means $\la E_x|E_x\ra= \la E_y|E_y\ra=\la E_z|E_z\ra\neq0$ and $\la E_i|E_j\ra=0$ with $i,j=x,y,z$ and $i\neq j$. Therefore the degree of 3D polarization coherence is defined \cite{Setala2002PRE,Barakat1977OC, Setala2002PRL} as 
\beq \label{3DP}
{\cal P}_3=\sqrt{\frac{3}{2}\left({\rm Tr}{\cal W}_{\rm 3D}^2-\frac{1}{3}\right)},
\eeq
which varies between 0 and 1 with 0 meaning complete unpolarization and 1 indicating fully polarized. Here, the subscript indicates the dimensionality 3. This measure is consistent with the conventional 2D definition of degree of polarization for arbitrary light beams \cite{BornWolf}, which will also be systematically discussed in the following. It means how much the light field is concentrated to a single polarization direction. Mathematically, it can be re-expressed through the eigenvalues $m_1$, $m_2$, $m_3$ of the normalized coherence matrix (\ref{3Dnorm-matrix}) as
\beqa \label{3DPeigen}
{\cal P}_3=\sqrt{1-\frac{2\times 3 (m_1m_2+m_1m_3+m_2m_3)}{3-1}}.
\eeqa

On the other hand, another coherence quantity, entanglement, between the polarization space $\{|x\ra,|y\ra,|z\ra\}$ and the amplitude space $\{|E_x\ra,|E_y\ra,|E_z\ra\}$ of the general 3D field represents a $3\times3$ bipartite pure-state scenario. Therefore, Schmidt analysis \cite{NC2000, Grobe1994JPB, Ekert1995AJP} can be applied with the quantitative Schmidt number measure $K=1/\sum^3_{i=1}\lambda_i^2$. Here $\sqrt{\lambda_i}$, $i=1,2,3$ are the Schmidt coefficients and can be shown to coincide with the eigenvalues of the normalized polarization coherence matrix (\ref{3Dnorm-matrix}), i.e., $\lambda_i=m_i$, $i=1,2,3$. The Schmidt number $K$ varies between $1$ and $3$ for the 3D light field, i.e., $K_3 \in [1,3]$, where $K_3=1$ indicates zero entanglement with only one nonzero Schmidt coefficient and $K_3=3$ means maximal entanglement with equal Schmidt coefficients $m_1=m_2=m_3$. 

To compare with the normalized 3D degree of polarization (\ref{3DP}), entanglement $K$ is also normalized as
\beq \label{3DKnorm}
{\cal K}_3=\sqrt{\frac{3}{2}\left(1-\frac{1}{K_3 }\right)}.
\eeq
Obviously, $\mathcal{K}_3$, called Schmidt weight \cite{Qian2018NJP}, $\in [0,1]$ with $\mathcal{K}_3 = 0,1$ meaning minimum (zero), maximal entanglement respectively. Some tedious but straightforward calculations show that $\mathcal{K}_3$ can be further expressed in terms of the eigenvalues
\beq \label{3DKeigen}
{\cal K}_3=\sqrt{3(m_1m_2+m_1m_3+m_2m_3)}. 
\eeq

By comparing equations (\ref{3DPeigen}) and (\ref{3DKeigen}), one can immediately arrive at the complementary identity relation 
\beq \label{3DPK}
{\cal P}_3^2 + \mathcal{K}_3^2 = 1.
\eeq
This is our first major result. It illustrates an intrinsic complementary behavior of polarization coherence with entanglement coherence for arbitrary light fields. It will be shown later that this result can reduce systematically to arbitrary two dimensional light beams and can also generalize to any N-dimensional structural fields. \\

\begin{figure}[h]
\centering
\includegraphics[width=5.5cm]{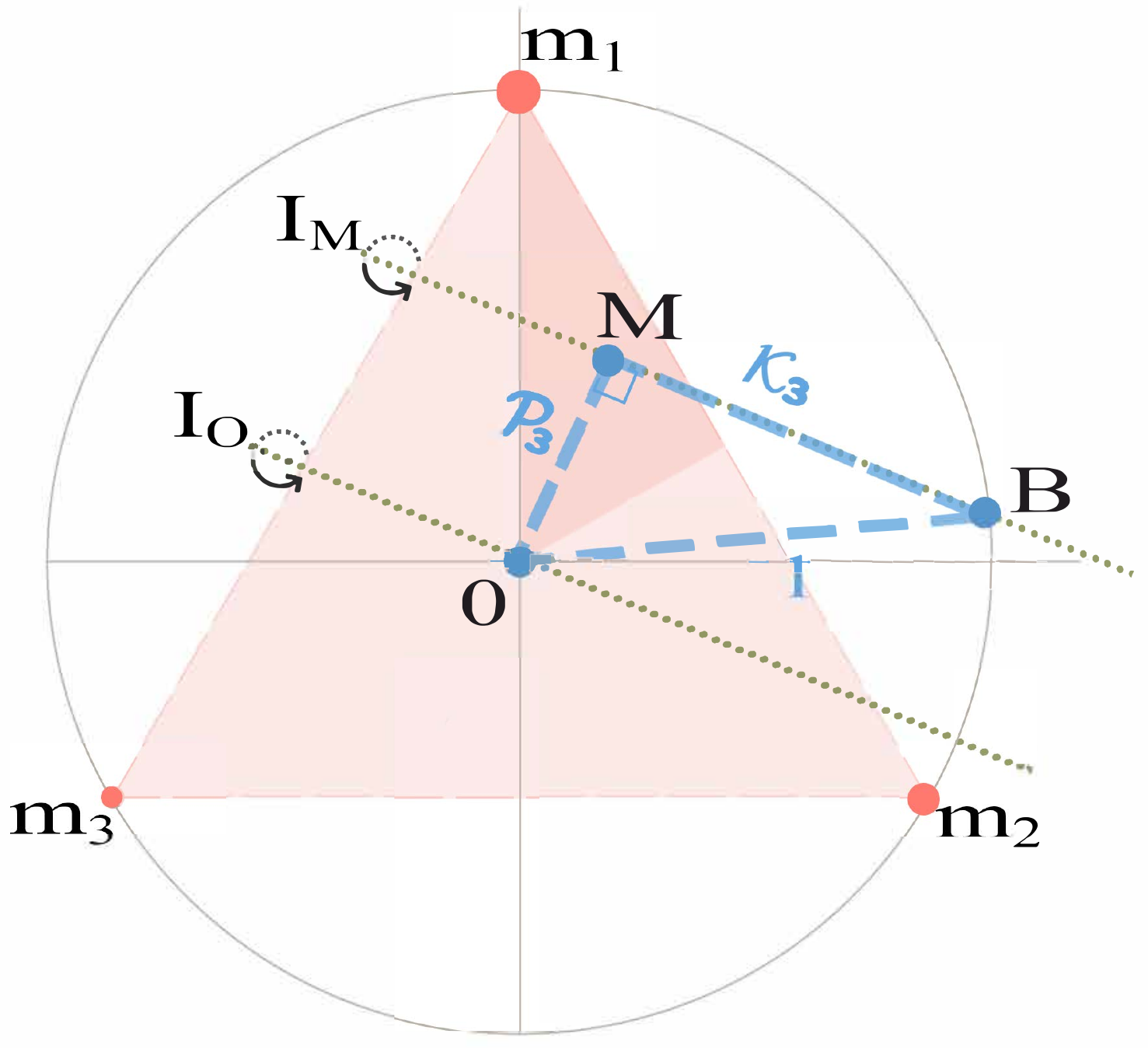}
\caption{Geometric illustration of mapping optical polarization coherence and entanglement to mechanical concepts of COM and MOI for arbitrary 3D light fields. The three masses $m_{1,2,3}$ are placed at the vertices of an equilateral triangle (that is inscribed in the circle $O$) and $M$ is the center of mass point. The lengths of $\overline{OM}$,  $\overline{MB}$ represent the values of degree of polarization ${\cal P}_3$, entanglement ${\cal K}_3$ respectively. Here the line $\overline{MB}$ is perpendicular to $\overline{OM}$ and it intersects with the circle $O$ at point $B$. The Pythagorean theorem of the right triangle $\triangle OMB$ represents directly the complementary relation (\ref{3DPK}). $I_M$, $I_O$ are the moment of inertia correspond to rotations along the entanglement line $\overline{MB}$ and its parallel partner that passes through point $O$ respectively. The sizes of the point-mass dots indicate $m_1\ge m_2\ge m_3$ without loss of any generality.}
\label{3D-COM}
\end{figure}


{\noindent \bf Center of Mass}: To further understand the optical coherence quantities and the above general complementary relation (\ref{3DPK}), we now describe a two-step geometric mapping procedure that links to mechanical concepts. {\em Step 1}: let the polarization coherence matrix eigenvalues $m_1$, $m_2$, $m_3$ represent the values of three point masses. {\em Step 2}: place these point masses at the vertices of an equilateral triangle inscribed in a unit circle $O$, see Fig.~\ref{3D-COM} for illustration. With such a mapping, it is then ready to analyze the connection to mechanical properties. 

The three-mass system has a center of mass point $M$ that is located inside the two-dimensional triangle $\triangle m_1m_2m_3$. Then the coordinates $(X^{(1)}, X^{(2)})$ of $M$ are simply determined as 
\beq
X^{(j)}m_{tot}=x_1^{(j)}m_1+x_2^{(j)}m_2+x_3^{(j)}m_3,
\eeq 
where $m_{tot}$ is the total mass and $x_i^{(j)}$ represents the $j$-th coordinate of the $i$-th mass $m_i$ with $i=1,2,3$ and $j=1,2$. When taking $O$ as the origin of the 2D coordinate system, the distance between $O$ and $M$ can be simply determined as $\overline{OM}=\sqrt{(X^{(1)})^2+(X^{(2)})^2}$. Furthermore, the distance between the mass center $M$ and point $B$, which is the cross point of the circle $O$ with line $\overline{MB}$ (perpendicular to $\overline{OM}$), can also be obtained directly as $\overline{MB}=\sqrt{1-(X^{(1)})^2-(X^{(2)})^2}$. Surprisingly, it can then be shown that
\beq \label{PK-COM}
{\cal P}_3=\overline{OM}, \quad \text {and} \quad {\cal K}_3=\overline{MB}.
\eeq 
That is, the degree of 3D polarization ${\cal P}_3$ equals the value of the distance between the geometric center $O$ and the mass center $M$, and the degree of entanglement ${\cal K}_3$ equals the value of the distance between mass center $M$ and point $B$. When the three masses are equal, the center of mass point $M$ coincides with the geometric center $O$ so that $\overline{OM}=0$ ($\overline{MB}=1$) indicating complete unpolarization ${\cal P}_3=0$ (maximal entanglement ${\cal K}_3=1$). When the total mass is concentrated on one point mass (e.g., $m_1$), the remaining two masses vanish. Then the center of mass $M$ coincides with point $m_1$ with $\overline{OM}=1$ ($\overline{MB}$=0) indicating complete polarization ${\cal P}_3=1$ (minimum entanglement ${\cal K}_3=0$). The detailed proof of this quantitative connection for the generalized N-dimensional case is given in Supplemental Material section A.

As a result, the polarization-entanglement complementary relation (\ref{3DPK}) can now be represented by the Pythagorean theorem of the right triangle $\triangle OMB$ that connects the mass center $M$, geometric center $O$, and point $B$, illustrated by the blue dashed lines in Fig.~\ref{3D-COM},
\beq \label{3DIdentity-COM}
{\cal P}_3^2 + \mathcal{K}_3^2 = 1 \quad \Leftrightarrow \quad \overline{OM}^2+\overline{MB}^2=\overline{OB}^2.
\eeq

Eqns.~(\ref{PK-COM}), (\ref{3DIdentity-COM}) represent the second major result showing direct quantitative connections of optical polarization, entanglement coherence, and their complementary relation to the mechanical concept of center of mass. \\


{\noindent \bf Moment of Inertia}: The center of mass of a system is related to another mechanical concept, moment of inertia, when the rotation axis passes through the mass center $M$. Combined with the above discussion, the entanglement line $\overline{OM}$, as shown in Fig.~\ref{3D-COM}, serves as a crucial rotation axis, about which the moment of inertia $I_M$ of the three-mass system can be obtained as $I_{M}=m_1r_1^2+m_2r_2^2+m_3r_3^2$, where $r_i$, $i=1,2,3$, are the distances of mass $m_i$ to the axis $\overline{OM}$. 

The moment of inertia with respect to the parallel line that passes through the geometric center $O$ (see illustration in Fig.~\ref{3D-COM}), can also be achieved as $I_{O}=m_1s_1^2+m_2s_2^2+m_3s_3^2$, where $s_i$, $i=1,2,3$, are the distances of mass $m_i$ to the parallel axis. Then the Huygens-Steiner theorem \cite{Huygens-MOI, Steiner-MOI} (also called parallel axis theorem) reads
\beq 
I_{O}=I_M+m_{tot}d^2,
\eeq
where $d$ is the distance between the two paralleled axes. This leads straightforwardly to the quantitative relations
\beqa \label{3D-MOI}
{\cal P}_3=\sqrt{I_O- I_M}     \quad \text {and} \quad  {\cal K}_3=\sqrt{1-I_O+ I_M},  
\eeqa
which is the third major result of the Letter. They establish direct quantitative connections between optical coherence quantities and mechanical quantities. 

The polarization coherence and entanglement of a generic light field can now be interpreted as the difference between two moment of inertia $I_O$ and $I_M$ of the three-mass system. Complete unpolarization ${\cal P}_3=0$ (or maximal entanglement ${\cal K}_3=1$) now simply means the moment of inertia $I_M$ coincides with $I_O$ so that $I_O- I_M=0$, while complete polarization ${\cal P}_3=1$ (or zero entanglement ${\cal K}_3=0$) indicates the moment of inertia $I_M$ and $I_O$ are maximally separated with $I_O- I_M=1$. This provides a new way of understanding and obtaining optical coherence quantities through the mapped point-mass scenario.

On the other hand, mechanical properties of such a three-mass system (or N-mass system as extended in the following) can also be understood and achieved with the optical polarization coherence and entanglement. These mechanical properties include center of mass, momentum of inertia, as well as their related properties such as angular momentum $L=I\omega$ with $\omega$ being rotation frequency, rotational energy $E=I\omega^2/2$, etc. \\


{\noindent \bf Generalization to arbitrary dimensions}: The above polarization-entanglement complementary relation (\ref{3DPK}) along with its connection to the mechanical concepts of center of mass and moment of inertia can be further generalized to arbitrary dimensional tensor structures. In this case, the concept of polarization is not restricted to describe the 3 wave oscillation directions of the 3D space anymore. It is extended to represent all vectors of a generic vector space of arbitrary dimension \cite{Qian2016PRL, Eberly2017O}. 

A generic $N$-dimensional two-space (or two-degree-of-freedom) structure can be written as 
\beq \label{generic form}
|E\ra=\sum^{N}_{l=1}|G_l\ra |Z_{l}\ra,
\eeq
where $|G_l\ra$ are normalized basis vectors of one vector space (e.g., the infinite dimensional spatial degree of freedom of light) and $|Z_{l}\ra$ represent the amplitudes that group all remaining degrees of freedom as a single large vector space (e.g., the combination of wave oscillation directions and temporal modes of light). 

The above state is a direct $N$-dimensional extension of the light field (\ref{3Dnorm-field}) but with the vector space $\{|G_l\ra\}$ singled out. The extended concept of ``polarization" simply means all the basis vectors $|G_l\ra$ (correspond $|x\ra, |y\ra, |z\ra$ in the general 3D light field case). Then the degree of ``polarization" is directly extended to mean how much this field $|E\ra$ is concentrated to a single superposed vector in this $G$ space. As a result, this generalized ``polarization" coherence can be systematically defined as \cite{Qian2016PRL, Eberly2017O}
\beq \label{NDP}
{\cal P}_N=\sqrt{\frac{N}{N-1}\left({\rm Tr}{\cal W}_{\rm ND}^2-\frac{1}{N}\right)},
\eeq
which is a direct extension of (\ref{3DP}) with 
\beq \label{NDmatrix}
{\cal W}_{ND}=\frac{\sum_{k,l}\la Z_k|Z_l\ra|G_l\ra\la G_k|}{\sum_{k}\la Z_k|Z_k\ra}
\eeq 
being the normalized $N$-dimensional (ND) ``polarization" coherence matrix of the $G$ space with $k,l=1,2,3,...,N$. Here, ${\cal P}_N$ is normalized between 0 and 1 indicating complete un-``polarization" and ``polarization" respectively.

Entanglement of the $N$-dimensional two-space field (\ref{generic form}) can be analyzed systematically by Schmidt decomposition as in the 3D case, and quantitatively measured by the Schmidt weight as
\beq \label{NDK}
{\cal K}_N=\sqrt{\frac{N}{N-1}\left(1-\frac{1}{K_N }\right)},
\eeq
where $K_N=1/\sum^N_{i=1}m_i^2$ is the Schmidt number with $\sqrt{m_i}$ being the Schmidt coefficients and the $m_i$ are also the eigenvalues of the coherence matrix ${\cal W}_{ND}$. Here ${\cal K}_N$ is bounded between 0 and 1 indicating zero and maximal entanglement respectively. 

Combining the degree of generic ``polarization" and entanglement for the generic ND structure, one is then led to the generalized identity 
\beq \label{NDPK}
{\cal P}_N^2+\mathcal{K}_N^2 = 1.
\eeq
This generic entanglement-polarization complementary relation suggests that the intrinsic opposite behaviors of ``polarization" and entanglement in a general field are universal. The absence of ``polarization" coherence is always accompanied by the display of maximal entanglement, and vice versa. The proof of (\ref{NDPK}) is provided in the supplemental material. 

\begin{figure}[t]
\centering
\includegraphics[width=5cm]{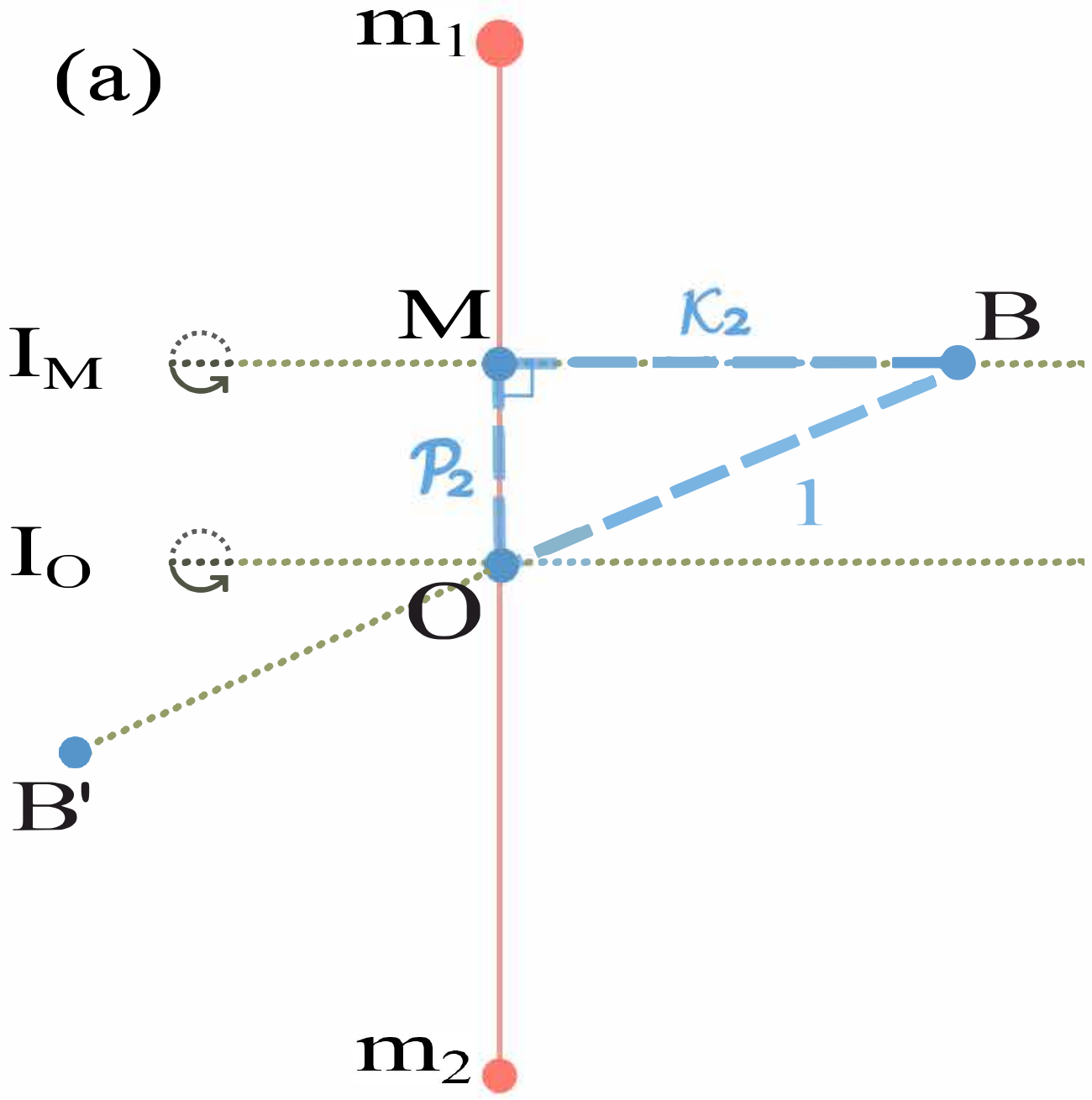}
\vspace{3mm}
\vspace{3mm}
\includegraphics[width=5.5cm]{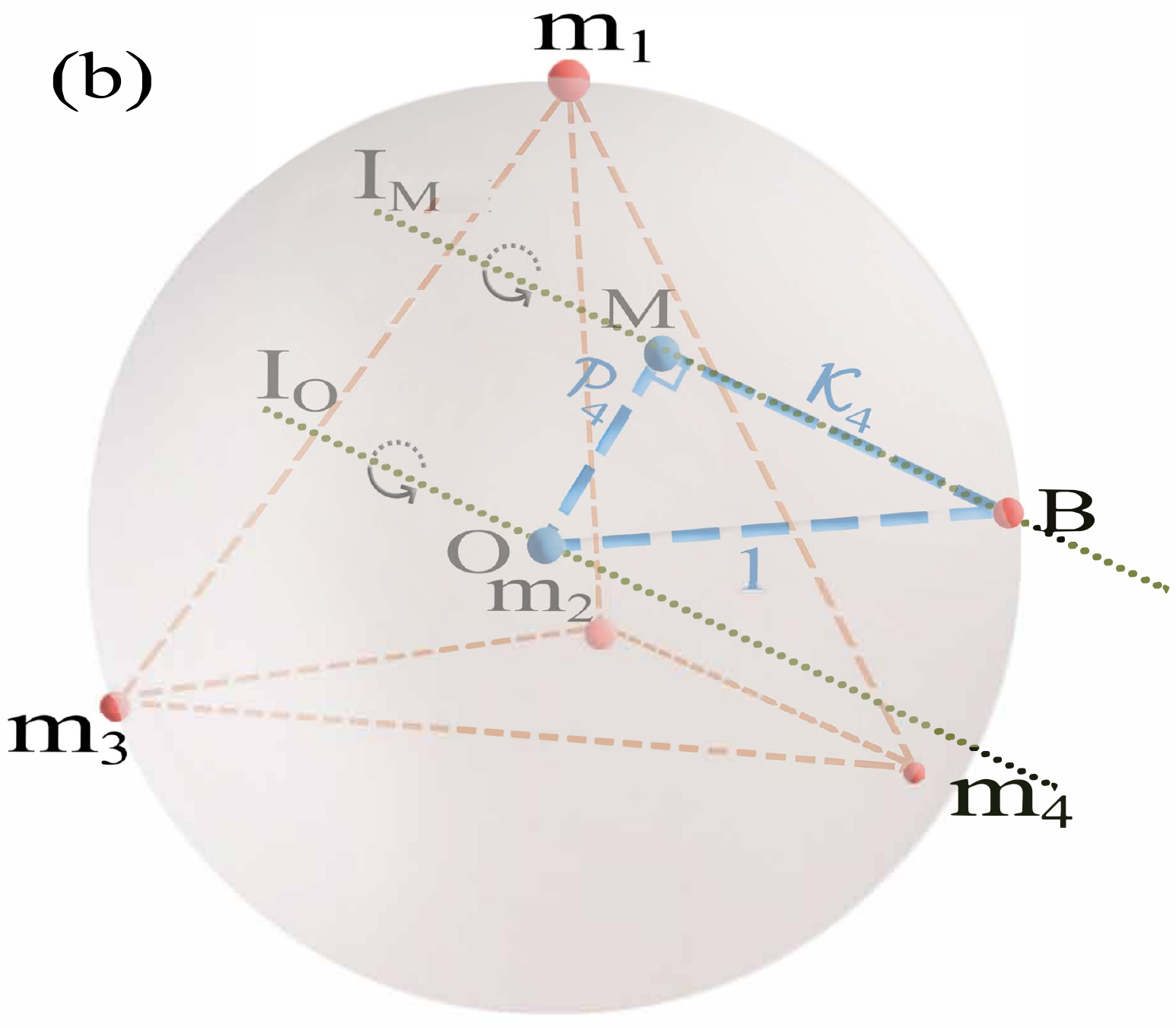}
\caption{Geometric illustrations of mapping optical polarization coherence and entanglement to mechanical concepts of COM and MOI for arbitrary 2D beams and 4D generic structural fields. For the 2D case in panel (a), two masses $m_{1,2}$ are placed at the vertices of the regular 1-simplex (a segment) and the 0-sphere $O$ has two end points $B, B'$. For the 4D case in panel (b), four masses $m_{1,2,3,4}$ are placed at the vertices of the regular 3-simplex (a tetrahedron) inscribed in the 2-sphere $O$. For both cases, $M$ is the center of mass and the lengths of $\overline{OM}$,  $\overline{MB}$ represent the values of degree of polarization ${\cal P}_N$, entanglement ${\cal K}_N$ respectively. Line $\overline{MB}$ is perpendicular to $\overline{OM}$ and it intersects with the ($N-2$)-sphere $O$ at point $B$. The Pythagorean theorem of the right triangle $\triangle OMB$ represents directly the complementary relation (\ref{NDPK}). $I_M$, $I_O$ are the moment of inertia correspond to rotations along the entanglement line $\overline{MB}$ and its parallel partner that passes through $O$ respectively. The sizes of the point-mass dots indicate $m_1\ge m_2\ge m_3 \ge m_4$ without loss of any generality.}
\label{figND}
\end{figure}

It is important to note that when $N=2$, the field reduces to the arbitrary two-dimensional optical beam case. Then ${\cal P}_2$ is exactly the conventional degree of polarization \cite{BornWolf,Wolf1959} and entanglement ${\cal K}_2$ is exactly the well-known entanglement measure concurrence \cite{Wootters1998PRL}. Fig.~\ref{figND} (a) illustrates of the 2D complementary relation with triangle $\triangle OMB$. This two-dimensional relation is consistent with wave-particle complementarity relations, see for example, in Refs.~\cite{Qian2016PRL, Norrman2017PRL, Qian2018O, Galazo2020PLA, Paul2020OL, Zela2020QR, Qureshi2021OL}.

The connection to mechanical concepts can also be extended systematically following the two-step geometric mapping prescription. {\em Step 1}: let the eigenvalues (or the square of the Schmidt coefficients) $m_1, ..., m_N$ of the ``polarization" coherence matrix (\ref{NDmatrix}) to represent the values of $N$ point masses. {\em Step 2}: place these point masses at the $N$ vertices of a regular ($N-1$)-simplex inscribed in a unit ($N-2$)-sphere with origin $O$. 

Consistent with the case of 3D generic light field, the value of degree of ``polarization" ${\cal P}_N$ is exactly the distance from $O$ to the center of mass point $M$, i.e., ${\cal P}_N=\overline{OM}$, and the value of degree of entanglement $\mathcal{K}_N=\overline{MB}$ where $\overline{MB}\perp\overline{OM}$ and $B$ is the cross point with the unit $(N-2)$-sphere $O$, see illustration in Fig.~\ref{figND}. Then the right triangle $\triangle OMB$ with $\overline{OM}^2+\overline{MB}^2=\overline{OB}^2$ directly represents the generic polarization-entanglement complementary relation ${\cal P}_N^2 + \mathcal{K}_N^2 = 1$. A detailed proof of these generalized results provided in the supplemental material. 

Furthermore, the moment of inertia of the $N$-mass system with respect to the entanglement line $\overline{MB}$ and the parallel line that passes through $O$ obey exactly the same quantitative relation as in the 3D case (\ref{3D-MOI}), connecting to optical ``polarization" coherence and entanglement as 
\beqa \label{ND-MOI}
{\cal P}_N=\sqrt{I_O- I_M}    \quad \text {and} \quad  {\cal K}_N=\sqrt{1-I_O+ I_M}.
\eeqa

To this end, we have shown that all three major results (\ref{3DPK}), (\ref{3DIdentity-COM}), (\ref{3D-MOI}) about the generic 3D light field can be reduced to arbitrary 2D beams and extended to arbitrary $N$D structural fields. The optical polarization-entanglement complementary relation is a universal feature for all light. The quantitative connections of optical coherence quantities (polarization and entanglement) with mechanical concepts of center of mass and moment of inertia are also universal for all light fields. \\


{\noindent \bf Summary}: In summary, we have established a universal polarization-entanglement complementary relation ${\cal P}_N^2+\mathcal{K}_N^2 = 1$ for arbitrary light fields of 2D and 3D polarizations and for general fields of $N$-dimensional structural ``polarization". The complementarity suggests that polarization and entanglement are indeed two intrinsically opposite coherence properties of all light fields. The absence of polarization coherence is always accompanied by the display of entanglement, vice versa.

A geometric mapping technique is introduced to correspond the eigenvalues of the ``polarization" coherence matrix (or the Schmidt coefficients) to point masses that are located at the vertices of a regular simplex. Based on the mapping, optical coherence quantities of polarization and entanglement (indication of the Huygens-Fresnel wave theory) are shown to be quantitatively connected to the seemingly unrelated mechanical concepts of center of mass and moment of inertia (result of the Huygens-Steiner theorem). 

The obtained quantitative relations in (\ref{PK-COM}), (\ref{3DIdentity-COM}), (\ref{3D-MOI}) and their $N$-dimensional extensions open a unique avenue to link coherence optics to mechanics. These relations provide a new way to interpret and understand the meaning of coherence optics concepts such as complete polarization, partial polarization, complete unpolarization, separable, partial entanglement, maximal entanglement, etc. They also help to quantitatively obtain and analyze coherence optical quantities with mechanical scenarios. On the other hand, these quantitative results also establishes a new platform to understand and obtain mechanical concepts of point-mass systems via coherence optical contexts. 

We expect many other related concepts of light such as coherence, correlation and entropy to be analyzed within the mechanical scopes, and additional relevant mechanical concepts e.g., angular momentum, rotational energy, etc., to be connected to optical coherence properties. 

Finally, the tensor structure of the generic light field (\ref{generic form}) is similar to that of a quantum pure state. Therefore our analysis of generalized ``polarization" and entanglement also applies to quantum states. \\

{ \noindent \bf Acknowledgment:} We acknowledge partial support from the U.S. Army under Contact No. W15QKN-18-D-0040 and from Stevens Institute of Technology.





\end{document}